\documentstyle[12pt,axodraw]{article}
\makeatletter
\def\appendix{\par\clearpage
  \setcounter{section}{0}
  \setcounter{subsection}{0}
  \@addtoreset{equation}{section}
  \def\@sectname{Appendix~}
  \def\theequation{\thesection.\arabic{equation}}
  \def\thesection{\Alph{section}}}
\makeatother

\renewcommand{\theequation}{\thesection.\arabic{equation}}

\begin{document}
\begin{titlepage}
\hskip 11cm \vbox{ \hbox{BUDKERINP/2002-32} \hbox{July 2002}}

\vskip 0.3cm \centerline{\bf Radiative corrections to
the quark-gluon-Reggeized quark vertex in QCD$^{~\ast}$}

\vskip 0.8cm \centerline{  M.I.~Kotsky$^{a,c}$, L.N.~Lipatov$^b$,
A.~Principe$^c$ and M.I.~Vyazovsky$^d$} \vskip 0.3cm
\centerline{\sl $^a$ Budker Institute for Nuclear Physics, 630090
Novosibirsk, Russia} \centerline{\sl $^b$ St.Petersburg Nuclear
Physics Institute,} \centerline{\sl Gatchina, 188300
St.Petersburg, Russia} \centerline{\sl $^c$ Istituto Nazionale di
Fisica Nucleare, Gruppo collegato}\centerline{\sl di Cosenza,
Arcavacata di Rende, I-87036 Cosenza, Italy} \centerline{\sl $^d$
St.Petersburg State University, St.Petersburg, Russia} \vskip 1cm

\begin{abstract}
This paper is devoted to the calculation of quark-gluon-Reggeized quark
effective vertex in perturbative QCD in the next-to-leading order. The
case of QCD with massive quarks is considered. This vertex has a number
of applications, in particular, the result can be used for determination
of the next-to-leading correction to the massive Reggeized quark
trajectory.
\end{abstract}

\vfill
\hrule
\vskip 0.3cm
\noindent
$^{\ast}${\it Work supported in part by INTAS and in part by the
Russian Fund of Basic Researches.}
\vfill
\end{titlepage}

\section{Introduction}
\setcounter{equation}{0}

One of important and widely applied properties of QCD is the Reggeization
of elementary particles. Contrary to QED, where only the electron
Reggeizes~\cite{1} while the photon does not~\cite{2}, in QCD both quarks
and gluons are Reggeized~\cite{3},~\cite{4}. The gluon Reggeization is the
base of the BFKL approach to the description of so-called high-energy 
semihard
processes in QCD. Such processes have two (hard) well separated energy
scales so that one has to sum large energy logarithms in all orders of
perturbation theory. Derived originally~\cite{5} in the Leading Logarithmic
Approximation (LLA) the BFKL equation is presently known up to the
next-to-leading order accuracy~\cite{6}~-~\cite{8}.

Basing on the quark Reggeization instead of the gluon one, the
BFKL- like equation for amplitudes mediated by two (interacting)
Reggeized quarks in $t$- channel was derived in~\cite{4} in the
LLA. This equation could be obviously useful to study the
high-energy behaviour of amplitudes with meson quantum numbers in
$t$- channel. However, the LLA has a big disadvantage related to
that no scale dependencies can be fixed there as it has been
continuously pointed before (see Refs.~\cite{6}~-~\cite{8}, for
example). So, exactly as in the BFKL case, a reliable theoretical
description is impossible without knowledge of the structure of
the radiative corrections.

To study these radiative corrections, one has to know, in particular,
the interactions of Reggeized quarks with elementary particles in the
Next-to-leading Logarithmic Approximation (NLA). This paper is devoted
to the NLA calculation of quark-gluon-Reggeized quark effective vertex
in QCD with massive quarks. Let us note, that quite a lot of information
about Reggeized quarks in the NLA is already available in literature.
Among this there is the result of Ref.~\cite{9} for the same vertex we
consider but in the massless QCD case. Our calculation confirms the
correctness of the Ref.~\cite{9}. The other important results are
the two-loop Regge trajectory of the massless Reggeized quark~\cite{10}
and the quasi-multi-Regge amplitudes with quark exchanges in crossing
channels~\cite{11}.

In order to reach our aim we consider the one-loop amplitude
$q\bar q \to gg$ of the Fig.~1
\begin{figure}[t]
\begin{center}
\setlength{\unitlength}{0.35mm}
\begin{picture}(150,150)(0,0)
\ArrowLine(0,130)(75,130)
\Gluon(75,130)(147,130){3}{5}
\ArrowLine(75,20)(0,20)
\Gluon(75,20)(147,20){3}{5}
\COval(75,75)(65,25)(0){Black}{White}
\Text(37,140)[c]{$p_A$}
\Text(37,10)[c]{$-p_B$}
\Text(113,140)[c]{$p_{A^\prime}$}
\Text(113,10)[c]{$p_{B^\prime}$}
\ArrowLine(147,130)(150,130)
\ArrowLine(147,20)(150,20)
\end{picture} \\
{\sl Fig.1 Schematic representation of $q\bar q \to gg$ process.}
\end{center}
\end{figure}
with quark quantum numbers and positive signature in the $t$- channel
in the Regge kinematics
\begin{equation}\label{11}
- u = - (p_A - p_{B^\prime})^2 \approx s = (p_A + p_B)^2 \rightarrow
\infty~,\ \ \ t = q^2 = (p_A - p_{A^\prime})^{2} - \mbox{fixed}~.
\end{equation}
Due to the quark Reggeization, the above amplitude ${\cal A}$ can be
written as follows~\cite{4}
\begin{equation}\label{12}
-{\cal A} = \bar\Gamma_B(q,s_0)(\not q - m)^{-1}\frac{1}{2}\left[ \left(
\frac{-s}{s_0} \right)^{\hat\omega(q)} + \left( \frac{-u}{s_0} \right)^
{\hat\omega(q)} \right]\Gamma_A(q,s_0)~,
\end{equation}
with $\bar\Gamma_B$ and $\Gamma_A$ being the quark-gluon-Reggeized quark
(QGR) effective interaction vertices which we are interested in here.
The parameter $s_0$ is artificial and the amplitude~(\ref{12}) does not
depend on it. The other notations in the relation above are: $m$ for
the quark mass and $\hat\omega(q)$ for the Reggeized quark trajectory.
In the LLA it looks
$$
\hat\omega^{(1)}(q) = \frac{g^2C_F}{2\pi}(\not q -
m)\int\frac{d^{D-2}k_\perp}{(2\pi)^{D-2}}\frac{\not q_\perp  -
\not k_\perp + m}{k_\perp^2[(q - k)_\perp^2 - m^2]}
$$
\begin{equation}\label{13}
= - g^2C_F\frac{\Gamma(1-\epsilon)}{(4\pi)^{2+\epsilon}}
\frac{(\not q + m)\not q}{q^2}\left( m^2-q^2 \right)
\int_0^1\frac{dx}{\left[ (1-x)\left( m^2-xq^2 \right)
\right]^{1-\epsilon}}~,
\end{equation}
where $\Gamma(z)$ is the Euler gamma- function, $C_F$ is related
to the standard notation for the $SU(N)$ colour group Casimir
operator in the fundamental representation
\begin{equation}\label{14}
t^at^a = C_FI = \frac{N^2-1}{2N}I~,
\end{equation}
$g$ is the gauge coupling constant and the integration is carried out over
$(D-2)$- dimensional vector orthogonal to the initial particle momenta
plane. Throughout all this note we use the dimensional regularization
with the space-time dimension $D = 4 + 2\epsilon$.

To calculate the one-loop amplitude ${\cal A}$~(\ref{12}) we follow
the $t$- channel unitarity approach developed in Ref.~\cite{12} which
allows to considerably simplify the problem. So, at the one-loop level
we have to consider two $t$- channel intermediate states: quark-gluon
and quark ones. In the next section we consider the former contribution.

\section{The branch-point contribution}
\setcounter{equation}{0}

The quark-gluon $t$- channel discontinuity of the amplitude ${\cal A}$
is given by an ordinary Cutkosky cut of the contributing diagrams as
it is depictured in the Fig.~2.
\begin{figure}[t]
\begin{center}
\setlength{\unitlength}{0.35 mm}
\begin{picture}(150,150)(0,0)
\ArrowLine(0,130)(75,130)
\Gluon(75,130)(147,130){3}{5}
\ArrowLine(75,20)(0,20)
\Gluon(75,20)(147,20){3}{5}
\GlueArc(75,75)(47,90,270){3}{5}
\ArrowArcn(75,75)(47,90,0)
\ArrowArcn(75,75)(47,0,270)
\CCirc(75,130){15}{Black}{White}
\CCirc(75,20){15}{Black}{White}
\DashLine(0,75)(150,75){10}
\Text(37,140)[c]{$p_A$}
\Text(37,10)[c]{$-p_B$}
\Text(113,140)[c]{$p_{A^\prime}$}
\Text(113,10)[c]{$p_{B^\prime}$}
\Text(150,65)[c]{$p=q+k$}
\Text(15,65)[c]{$k$}
\ArrowLine(147,130)(150,130)
\ArrowLine(147,20)(150,20)
\ArrowLine(40,110)(42,112)
\ArrowLine(40,40)(38,42)
\end{picture} \\
{\sl Fig.2 Schematic representation of the $t$- channel quark-gluon
intermediate state.}
\end{center}
\end{figure}
After this cut we come to the consideration of the convolution of two
on-mass-shell Born amplitudes related to the upper ($A$) part of the
Fig.~2 and the lower one ($B$). Because of the on-mass-shellness and
since the external gluons are physical there is the invariance of the
$A$ and $B$ under gauge transformations of the intermediate gluon's
polarization and one is allowed to sum up over this polarization in an
arbitrary gauge. We choose the Feynman gauge to perform this sum, so that
we use $-g_{\mu\nu}$ for the polarization tensor. Then, after this
convolution one is allowed to perform the loop integration with the
complete propagators instead of the on-mass-shell $\delta$- functions.
The amplitude obtained in this way has the same $t$- channel singularities
as the complete one that is enough to restore the correct Regge asymptotics
according to the conclusions of the Ref.~\cite{12}. So we have to consider
\begin{equation}\label{21}
{\cal A} = \int\frac{d^Dp}{i(2\pi)^D}\frac{\sum(-)BA}{\left( k^2 + i\delta
\right)\left( p^2 - m^2 + i\delta \right)}~,\ \ \ k = p - q~,
\end{equation}
where the convolution is performed on-mass-shell and the notations for
relevant momenta are given in the Fig.~2.

The amplitude $A$ has a form
$$
A = -ig^2\bar u(p)\biggl\{ \left( t^ct^{A^\prime} \right)_{iA}\biggl[
\not e(\not q - m)^{-1}\not e^*_{A^\prime} - \frac{2}{s_1}\left( \not
e^*_{A^\prime}(ep_{A^\prime}) + \not e(e^*_{A^\prime}k) + (\not q - m)
(ee^*_{A^\prime}) \right) \biggr]
$$
$$
+ \left( t^{A^\prime}t^c \right)_{iA}\biggl[(\not q - m)\left( \frac
{\not e^*_{A^\prime}\not e}{u_1 - m^2} + \frac{2(ee^*_{A^\prime})}{s_1}
\right)
$$
\begin{equation}\label{22}
+ 2\not e^*_{A^\prime}\left( \frac{(ep_A)}{u_1-m^2} + \frac{(ep_{A^
\prime})}{s_1} \right) + 2\not e(e^*_{A^\prime}k)\left( \frac{1}{u_1-m^2}
+ \frac{1}{s_1} \right) \biggr] \biggr\}u_A~,
\end{equation}
and we have an analogous expression for the $B$. We introduce the notations
$e,\ c\ (u(p),\ i)$ for the intermediate gluon (quark) spin wave function
and colour index respectively. Other notations for the external particles
spin wave functions are evident and for the external gluons we use the
light-cone gauges
\begin{equation}\label{23}
(e_{A^\prime}p_{A^\prime}) = (e_{B^\prime}p_{B^\prime}) = (e_{A^\prime}
p_{B^\prime}) = (e_{B^\prime}p_{A^\prime}) = 0~,
\end{equation}
which mean, in other words, that the final result in general gauges will
be given by the replacements
\begin{equation}\label{24}
e_{A^\prime} \rightarrow e_{A^\prime} - \frac{p_{A^\prime}(e_{A^\prime}
p_{B^\prime})}{(p_{A^\prime}p_{B^\prime})}~,\ \ \ e_{B^\prime} \rightarrow
e_{B^\prime} - \frac{p_{B^\prime}(e_{B^\prime}p_{A^\prime})}{(p_{A^\prime}
p_{B^\prime})}~.
\end{equation}
We also introduce the intermediate invariants according to as follows
\begin{equation}\label{25}
u_1 = (k + p_A)^2~,\ \ \ s_1 = (k - p_{A^\prime})^2~,\ \ \ u_2 = (k -
p_B)^2~,\ \ \ s_2 = (k + p_{B^\prime})^2~.
\end{equation}

Looking at the expression of the Eq.~(\ref{22}) for the amplitude $A$, one
can realize that its convolution with the $B$ is simple, but rather long,
nevertheless. For this reason we explain simplifications we use on a
relatively simple example of the convolution of so-called ``asymptotic''
parts~\cite{12} of the $A$ and $B$ which we choose in the form
$$
A_{as} = -ig^2\left( t^ct^{A^\prime} \right)_{iA}\bar u(p)\left(
\not e + (\not q - m)\left[ \frac{(ep_A)}{u_1-m^2} - \frac{(ep_
{A^\prime})}{s_1} \right] \right)(\not q - m)^{-1}\not e^*_{A^
\prime}u_A~,
$$
\begin{equation}\label{26}
B_{as} = -ig^2\left( t^{B^\prime}t^c \right)_{Bi}\bar v_B\not e^*_
{B^\prime}(\not q - m)^{-1}\left( \not e^* - (\not q - m)\left[
\frac{(e^*p_B)}{u_2-m^2} - \frac{(e^*p_{B^\prime})}{s_2} \right]
\right)u(p)~.
\end{equation}
This terminology ``asymptotic'' is because the amplitudes $A_{as}$
and $B_{as}$ give the asymptotics of the complete amplitudes $A$
and $B$ in their Regge limits $|u_1| \approx |s_1| \gg |t|$ and
$|u_2| \approx |s_2| \gg |t|$, respectively. They are invariant
under gauge transformations of the intermediate gluon's polarization
as well as the complete amplitudes (let us remind that the
prescription~(\ref{24}) is supposed in the relations~(\ref{26})).

The convolution of the asymptotic parts has a form
$$
\sum(-)B_{as}A_{as} = - g^4C_F\left( t^{B^\prime}t^{A^\prime} \right)
_{BA}\bar v_B\biggl\{ \not e^*_{B^\prime}(\not q - m)^{-1}\gamma^\mu
(\not p + m)\gamma_\mu(\not q - m)^{-1}\not e^*_{A^\prime}
$$
$$
+ \not e^*_{B^\prime}(\not q - m)^{-1}\biggl[ \frac{\not p_A}{u_1-m^2} -
\frac{\not p_{A^\prime}}{s_1} \biggr](\not p + m)\not e^*_{A^\prime} -
\not e^*_{B^\prime}(\not p + m)\biggl[ \frac{\not p_B}{u_2-m^2} - \frac
{\not p_{B^\prime}}{s_2} \biggr](\not q - m)^{-1}\not e^*_{A^\prime}
$$
\begin{equation}\label{27}
- \not e^*_{B^\prime}(\not p + m)\not e^*_{A^\prime}\frac{s}{2}\biggl[
\frac{1}{(u_2-m^2)(u_1-m^2)} + \frac{1}{s_2s_1} - \frac{1}{(u_2-m^2)s_1}
- \frac{1}{s_2(u_1-m^2)} \biggr] \biggr\}u_A~.
\end{equation}
When the first term in the curly brackets of the above relation is put into
the Eq.~(\ref{21}) as an integrand, the result of integration can be 
expressed
through the vector $q$ only. Therefore, performing the projection we 
conclude
that the integration momentum $p$ in this term can be replaced by
\begin{equation}\label{28}
p \rightarrow q\frac{m^2+q^2}{2q^2}~,
\end{equation}
where we have applied the on-mass-shellness of the intermediate particles
too. Therefore we obtain
$$
\not e^*_{B^\prime}(\not q - m)^{-1}\gamma^\mu(\not p + m)\gamma_\mu
(\not q - m)^{-1}\not e^*_{A^\prime}
$$
\begin{equation}\label{29}
= - \not e^*_{B^\prime}(\not q - m)^{-1}\left( (1+\epsilon)(\not q - m)
\frac{\not q}{q^2}(\not q - m) - 2m \right)(\not q - m)^{-1}\not 
e^*_{A^\prime}
\end{equation}
for the first term of the Eq.~(\ref{27}). Analogously, the integration 
momentum
in the numerator of the second term of the Eq.~(\ref{27}) is expressed in 
terms
of the vectors $p_{A^\prime}$ and $q$ (see Eq.~(\ref{25})). The projection
gives
\begin{equation}\label{210}
p \rightarrow \frac{1}{m^2-q^2}\left[ \frac{(m^2-q^2)(m^2+q^2) - 2(u_1
-m^2)q^2}{m^2-q^2}p_{A^\prime} + (u_1-m^2)q \right]~,
\end{equation}
and this term gets a form
$$
\not e^*_{B^\prime}(\not q - m)^{-1}\biggl[ \frac{\not p_A}{u_1-m^2} -
\frac{\not p_{A^\prime}}{s_1} \biggr](\not p + m)\not e^*_{A^\prime} =
\not e^*_{B^\prime}(\not q - m)^{-1}\biggl\{ (2e^*_{A^\prime}q)\biggl[
\left( \frac{2m^2\not q}{m^2-q^2} + m - \not q \right)
$$
$$
\times\frac{1}{u_1-m^2} + \frac{\not q-m}{s_1} - \frac{2m^2\not q}{(m^2
-q^2)^2} \biggr]
$$
\begin{equation}\label{211}
- \biggl[ \frac{q^2+2m^3(\not q-m)^{-1}}{m^2-q^2} + \left( m\not q + m^2
- q^2 - 2mq^2(\not q-m)^{-1} \right)\frac{1}{u_1-m^2} \biggr]\not e^*
_{A^\prime} \biggr\}~.
\end{equation}
The consideration of the third term in the curly brackets of the
Eq.~(\ref{27}) repeats the previous one with evident changes. The
corresponding relations are
$$
p \rightarrow \frac{1}{m^2-q^2}\left[ - \frac{(m^2-q^2)(m^2+q^2) - 2(u_2
-m^2)q^2}{m^2-q^2}p_{B^\prime} + (u_2-m^2)q \right]~,
$$
$$
\not e^*_{B^\prime}(\not p + m)\biggl[ \frac{\not p_B}{u_2-m^2} - \frac
{\not p_{B^\prime}}{s_2} \biggr](\not q - m)^{-1}\not e^*_{A^\prime} =
- \biggl\{ (2e^*_{B^\prime}q)\biggl[ \left( \frac{2m^2\not q}{m^2-q^2} +
m - \not q \right)\frac{1}{u_2-m^2}
$$
$$
+ \frac{\not q-m}{s_2} - \frac{2m^2\not q}{(m^2-q^2)^2} \biggr] - \not e
^*_{B^\prime}\biggl[ \frac{q^2+2m^3(\not q-m)^{-1}}{m^2-q^2}
$$
\begin{equation}\label{212}
+ \left( m\not q + m^2 - q^2 - 2mq^2(\not q-m)^{-1} \right)\frac{1}{u_2-
m^2} \biggr] \biggr\}(\not q-m)^{-1}\not e^*_{A^\prime}~.
\end{equation}
Finally, the integration momentum $p$ in the numerator of the last term
of the Eq.~(\ref{27}) is expressed through the complete basis of the problem
$p_{A^\prime}$, $p_{B^\prime}$ and $q$
\begin{equation}\label{213}
p \rightarrow \approx \frac{u_1-m^2}{s}\left( \frac{m^2-q^2}{2q^2}q +
p_{B^\prime} \right) + \frac{u_2-m^2}{s}\left( \frac{m^2-q^2}{2q^2}q -
p_{A^\prime} \right) + \frac{m^2+q^2}{2q^2}q~,
\end{equation}
where we have kept the terms which survive (after the integration in the
Eq.~(\ref{21})) in the Regge limit~(\ref{11}) only. Taking into account the
previous relation we get
$$
\not e^*_{B^\prime}(\not p + m)\not e^*_{A^\prime}\frac{s}{2}\biggl[
\frac{1}{(u_2-m^2)(u_1-m^2)} + \frac{1}{s_2s_1} - \frac{1}{(u_2-m^2)s_1}
- \frac{1}{s_2(u_1-m^2)} \biggr]
$$
$$
= - \not e^*_{B^\prime}(\not q-m)^{-1}\left( \frac{1}{u_1-m^2} - \frac{1}
{s_1} \right)\left( (\not q-m)(2e^*_{A^\prime}q) + \frac{m^2-q^2}{2q^2}
(q^2+m\not q)\not e^*_{A^\prime} \right)
$$
$$
- \left( (2e^*_{B^\prime}q)(\not q-m) + \not e^*_{B^\prime}(q^2+m\not q)
\frac{m^2-q^2}{2q^2} \right)\left( \frac{1}{u_2-m^2} - \frac{1}{s_2}
\right)(\not q-m)^{-1}\not e^*_{A^\prime}
$$
\begin{equation}\label{214}
- \not e^*_{B^\prime}(\not q-m)^{-1}\frac{m^2-q^2}{4q^2}(q^2+m\not q)s
\left( \frac{1}{u_2-m^2} - \frac{1}{s_2} \right)\left( \frac{1}{u_1-m^2}
- \frac{1}{s_1} \right)\not e^*_{A^\prime}~.
\end{equation}
Combining the relations~(\ref{29},~\ref{211},~\ref{212},~\ref{214})
according to the Eq.~(\ref{27}) we come to the following equality
$$
\sum(-)B_{as}A_{as} = - g^4C_F\left( t^{B^\prime}t^{A^\prime} \right)_
{BA}\bar v_B\biggl( \not e^*_{B^\prime}(\not q-m)^{-1}\biggl\{ \left(
\frac{m^2-q^2}{u_1-m^2} - 1 \right)\frac{2m^2\not q}{(m^2-q^2)^2}
(2e^*_{A^\prime}q)
$$
$$
+ \biggl[ (1+\epsilon)\frac{\not q(\not q+m)}{2q^2} - \epsilon + m(\not
q-m)^{-1} + m^2(\not q-m)^{-2} + \left( \not q + m^2(\not q-m)^{-1}
\right)\frac{\not q+m}{u_1-m^2}
$$
$$
+ \left( \frac{1}{u_1-m^2} - \frac{1}{s_1} \right)\frac{m^2-q^2}{2q^2}
\not q(\not q+m) \biggr]\not e^*_{A^\prime} \biggr\} + \biggl\{ (2e^*_
{B^\prime}q)\frac{2m^2\not q}{(m^2-q^2)^2}\left( \frac{m^2-q^2}{u_2-m^2}
- 1 \right)
$$
$$
+ \not e^*_{B^\prime}\biggl[ (1+\epsilon)\frac{(\not q+m)\not q}{2q^2}
- \epsilon + m(\not q-m)^{-1} + m^2(\not q-m)^{-2} + \frac{\not q+m}
{u_2-m^2}\left( \not q + m^2(\not q-m)^{-1} \right)
$$
$$
+ (\not q+m)\not q\frac{m^2-q^2}{2q^2}\left( \frac{1}{u_2-m^2} - \frac
{1}{s_2} \right) \biggr] \biggr\}(\not q-m)^{-1}\not e^*_{A^\prime}
$$
\begin{equation}\label{215}
+ \not e^*_{B^\prime}(\not q-m)^{-1}\frac{m^2-q^2}{4q^2}\not q(\not q+m)
s\left( \frac{1}{u_2-m^2} - \frac{1}{s_2} \right)\left( \frac{1}{u_1-m^2}
- \frac{1}{s_1} \right)\not e^*_{A^\prime} \biggr)u_A~.
\end{equation}
Let us note, that the Dirac equations as well as the conditions~(\ref{23})
were also used in order to express the convolution in the above form.
This convolution is to be put into the Eq.~(\ref{21}) as the integrand
and only scalar loop integrals appear there.

We see that in the $t$- channel unitarity approach the amplitude
${\cal A}$~(\ref{12}) is naturally expressed through a set of
independent scalar loop integrals. The algebra used for this purpose
is strongly simplified by the on-mass-shellness of the intermediate
particles and by the particular (Regge) kinematics we are interested
in. We have shown that for the simple contribution of the asymptotic
parts of the intermediate amplitudes $A$ and $B$ but evidently the
same can be applied to the complete amplitude ${\cal A}$. Here we
skip the details of the complete consideration and just quote the
result
$$
{\cal A}_2 = -g^2\left( t^{B^\prime}t^{A^\prime} \right)_{BA}\bar v
_B\not e^*_{B^\prime}(\not q-m)^{-1}\frac{\hat\omega(q)}{2}\left(
\ln\left( \frac{-s}{s_0} \right) + \ln\left( \frac{-u}{s_0} \right)
\right)\not e^*_{A^\prime}u_A
$$
$$
- g^4C_F\left( t^{B^\prime}t^{A^\prime} \right)_{BA}\bar v_B\biggl(
\not e^*_{B^\prime}(\not q-m)^{-1}\biggl\{ - \biggl[ m(\not q-m)^{-1}
\not e^*_{A^\prime} + \left( 2 - (1+\epsilon)\frac{(\not q-m)\not q}
{2q^2} \right)\not e^*_{A^\prime}
$$
$$
- \left( 2m + (1+\epsilon)\frac{m^2-q^2}{q^2}\not q + (1+2\epsilon)
(\not q-m) \right)\not e^*_{A^\prime}(\not q+m)^{-1} - \left( 1 +
\epsilon - \not q(\not q+m)^{-1} \right)
$$
$$
\times\frac{\not q\not e^*_{A^\prime}(\not q+m)^{-1}}{C_AC_F} \biggr]
\left( I_1 - I_1(q^2=m^2) \right) + m(\not q+m)\not e^*_{A^\prime}I_1
^\prime(q^2=m^2) + \frac{\not q-m}{C_AC_F}
$$
$$
\times\biggl[ \frac{m^2-q^2}{2q^2}\not q\not e^*_{A^\prime} - m\left(
(1-\epsilon)m-\epsilon\not q \right)\not e^*_{A^\prime}(\not q+m)^{-1}
\biggr]I_2 + \left( 2 + \frac{1}{C_AC_F} \right)\frac{m^2-q^2}{2q^2}
(\not q-m)
$$
$$
\times\not q\not e^*_{A^\prime}I_3 + \frac{m^2-q^2}{8q^2}(\not
q+m) \not q\not e^*_{A^\prime}\biggl[ \left( 2 + \frac{1}{C_AC_F}
\right)^2 I_4 + \frac{I_5}{C_A^2C_F^2} - \frac{2I_6}{C_AC_F}\left(
2 + \frac{1}{C_A C_F} \right) \biggr] \biggr\}
$$
$$
+ \biggl\{ - \biggl[ m\not e^*_{B^\prime}(\not q-m)^{-1} + \not e^*_{B^
\prime}\left( 2 - (1+\epsilon)\frac{\not q(\not q-m)}{2q^2} \right) -
(\not q+m)^{-1}\not e^*_{B^\prime}
$$
$$
\times\left( 2m + (1+\epsilon)\frac{m^2-q^2}{q^2}\not q + (1+2\epsilon)
(\not q-m) \right) - \frac{(\not q+m)^{-1}\not e^*_{B^\prime}\not q}
{C_AC_F}
$$
$$
\times\left( 1 + \epsilon - (\not q+m)^{-1}\not q \right) \biggr]\left(
I_1 - I_1(q^2=m^2) \right) + m\not e^*_{B^\prime}(\not q+m)I_1^\prime
(q^2=m^2)
$$
$$
+ \biggl[ \not e^*_{B^\prime}\not q\frac{m^2-q^2}{2q^2} - m(\not q+m)^
{-1}\not e^*_{B^\prime}\left( (1-\epsilon)m-\epsilon\not q \right) \biggr]
\frac{\not q-m}{C_AC_F}I_2 + \left( 2 + \frac{1}{C_AC_F} \right)\frac
{m^2-q^2}{2q^2}
$$
$$
\times\not e^*_{B^\prime}\not q(\not q-m)I_3 + \not e^*_{B^\prime}\not q
(\not q+m)\frac{m^2-q^2}{8q^2}
$$
\begin{equation}\label{216}
\times\biggl[ \left( 2 + \frac{1}{C_AC_F} \right)^2I_4 +
\frac{I_5}{C_A^2C_F ^2} - \frac{2I_6}{C_AC_F}\left( 2 +
\frac{1}{C_AC_F} \right) \biggr] \biggr\}(\not q-m)^{-1}\not
e^*_{A^\prime} \biggr)u_A~,
\end{equation}
where the large energy logarithms responsible for the quark Reggeization
were explicitly written in the first term of the ${\cal A}_2$.
The subscript $2$ in the notation ${\cal A}_2$
is to say that this part of the complete amplitude ${\cal A}$ has the
correct $t$- channel singularities related to the two-particle
$t$- channel intermediate state. As for the pole singularity contribution
appeared ``by accident'' at the above consideration, it was completely
removed from the ${\cal A}_2$ in order to be correctly restored from
the one-particle $t$-channel unitarity relation. That will be done in
the next section. Also, the amplitude ${\cal A}_2$ was already projected
on the positive signature and quark colour quantum numbers in the
$t$-channel since we are interested only in these quantum numbers as it
was mentioned above. The definitions of the six independent scalar
integrals entering the Eq.~(\ref{216}) are the
\begin{figure}[t]
\begin{center}
\hspace{-0.5cm}
\setlength{\unitlength}{0.35 mm}
\begin{picture}(450,150)(0,0)

\Line(20,20)(127,20)
\ArrowLine(127,20)(130,20)
\Gluon(75,20)(75,127){3}{5}
\ArrowLine(75,127)(75,130)
\Text(85,140)[c]{$k \to 0, c, \lambda$}
\Text(30,10)[c]{$q+k$}
\Text(130,10)[c]{$q$}

\Line(170,20)(277,20)
\ArrowLine(277,20)(280,20)
\Gluon(225,20)(225,127){3}{5}
\ArrowLine(225,127)(225,130)
\GlueArc(225,20)(32,180,360){3}{5}
\Text(235,140)[c]{$k \to 0, c, \lambda$}
\Text(175,10)[c]{$q+k$}
\Text(280,10)[c]{$q$}

\Line(320,20)(427,20)
\ArrowLine(427,20)(430,20)
\Gluon(375,52)(375,127){3}{5}
\ArrowLine(375,127)(375,130)
\GlueArc(375,20)(32,0,180){3}{5}
\Text(385,140)[c]{$k \to 0, c, \lambda$}
\Text(325,10)[c]{$q+k$}
\Text(430,10)[c]{$q$}

\end{picture} \\ \vspace{0.8cm}
{\sl Fig.3 The diagrams for a gluon emission by a quark.}
\end{center}
\end{figure}
following
$$
I_1 = \int\frac{d^Dp}{i(2\pi)^D}\frac{1}{\left( (p-q)^2 + i\delta \right)
\left( p^2 - m^2 + i\delta \right)}~,
$$
$$
I_2 = \int\frac{d^Dp}{i(2\pi)^D}\frac{1}{\left( (p-q)^2 + i\delta \right)
\left( p^2 - m^2 + i\delta \right)\left( (p+p_{A^\prime})^2 - m^2 + i\delta
\right)}~,
$$
$$
I_3 = \int\frac{d^Dp}{i(2\pi)^D}\frac{1}{\left( (p-q)^2 + i\delta \right)
\left( p^2 - m^2 + i\delta \right)\left( (p-p_A)^2 + i\delta \right)}~,
$$
$$
I_4 = \int\frac{d^Dp}{i(2\pi)^D}\frac{s}{\left( (p-q)^2 + i\delta \right)
\left( p^2 - m^2 + i\delta \right)\left( (p-p_A)^2 + i\delta \right)
\left( (p+p_B)^2 + i\delta \right)}
$$
$$
- \omega(q^2)\ln\left( \frac{-s}{s_0} \right)~,\ \ \ I_5 = \int\frac{d^Dp}
{i(2\pi)^D}
$$
$$
\times\frac{s}{\left( (p-q)^2 + i\delta \right)\left( p^2 - m^2 + i\delta
\right)\left( (p-p_{B^\prime})^2 - m^2 + i\delta \right)\left( (p+p_{A^
\prime})^2 - m^2 + i\delta \right)}
$$
$$
- \omega(q^2)\ln\left( \frac{-s}{s_0} \right)~,\ \ \ I_6 = \int\frac{d^Dp}
{i(2\pi)^D}
$$
\begin{equation}\label{217}
\times\frac{u}{\left( (p-q)^2 + i\delta \right)\left( p^2 - m^2 + i
\delta \right)\left( (p-p_{B^\prime})^2 - m^2 + i\delta \right)\left( (p-
p_A)^2 + i\delta \right)} - \omega(q^2)\ln\left( \frac{-u}{s_0} \right)~,
\end{equation}
with $\omega(q^2)$ defined by the Eq.~(\ref{13}) and
\begin{equation}\label{218}
\hat\omega(q) = g^2C_F\frac{m^2-q^2}{q^2}(\not q+m)\not q\omega(q^2)~.
\end{equation}
As it is clear from the LLA results of Ref.~\cite{4} the terms with
$\omega(q^2)$ cancel the large energy logarithms in the box integrals
$I_4$ - $I_6$. The results of integration in the Eqs.~(\ref{217}) are
listed in the Appendix of this paper.

\section{The pole contribution}
\setcounter{equation}{0}

The calculation of the pole contribution is much simpler than the
previous one. One has to convolute two amplitudes of a real gluon
emission by an on-mass-shell quark. Diagrammatically such the
amplitude is presented in the Fig.~3 where the diagrams related
to the external lines renormalization are not explicitly shown.
The calculation of this amplitude is done according to the ordinary
rules and is very simple due to the on-mass-shellness. Because of the
simplicity we just present below the result for the renormalized
amplitude skipping the details of this calculation
$$
A_{QQG}^\lambda = -ig\gamma^\lambda t^c\biggl( 1 + g^2C_F\biggl[ \frac
{\Gamma(1-\epsilon)}{(4\pi)^{2+\epsilon}}\frac{\sum_f\left( m_f^2
\right)^\epsilon}{3\epsilon C_F} + \biggl( 1 + \frac{1}{2C_AC_F} \biggr)
$$
\begin{equation}\label{31}
\times\biggl( \frac{\Gamma(1-\epsilon)}{(4\pi)^{2+\epsilon}}\frac{\left(
m^2 \right)^\epsilon}{\epsilon} + (3+2\epsilon)I_1(q^2=m^2) + 4m^2I_1^
\prime(q^2=m^2) \biggr) \biggr] \biggr)~,
\end{equation}
where the integral $I_1$ was defined by the Eqs.~(\ref{217}) and
the sum is over active quark flavours (including the quark with
the mass $m$ by which the gluon is emitted). The notation
$\gamma^\lambda$ in the above relation is for the Dirac matrix and
the other notations there are clear from the Fig.~3.

Now one has to perform the convolution of the two above on-mass-shell
amplitudes in the intermediate $t$- channel quark quantum numbers,
multiply it with the external particles wave functions and then,
exactly as in the previous section, replace the on-mass-shell $\delta$-
function of the intermediate quark by the complete propagator. Doing so
one gets the amplitude
$$
{\cal A}_1 = -g^2\left( t^{B^\prime}t^{A^\prime} \right)_{BA}\bar v_B
\biggl( \not e^*_{B^\prime}(\not q-m)^{-1}\not e^*_{A^\prime} + \not
e^*_{B^\prime}(\not q-m)^{-1}g^2C_F\biggl[ \frac{\Gamma(1-\epsilon)}
{(4\pi)^{2+\epsilon}}\frac{\sum_f\left( m_f^2 \right)^\epsilon}{3\epsilon
C_F}
$$
$$
+ \biggl( 1 + \frac{1}{2C_AC_F} \biggr)\biggl( \frac{\Gamma(1-\epsilon)}
{(4\pi)^{2+\epsilon}}\frac{\left( m^2 \right)^\epsilon}{\epsilon} + (3+2
\epsilon)I_1(q^2=m^2) + 4m^2I_1^\prime(q^2=m^2) \biggr) \biggr]\not e^*_
{A^\prime}
$$
$$
+ g^2C_F\not e^*_{B^\prime}\biggl[ \frac{\Gamma(1-\epsilon)}{(4\pi)^{2+
\epsilon}}\frac{\sum_f\left( m_f^2 \right)^\epsilon}{3\epsilon C_F} +
\biggl( 1 + \frac{1}{2C_AC_F} \biggr)
$$
\begin{equation}\label{32}
\times\biggl( \frac{\Gamma(1-\epsilon)}{(4\pi)^{2+\epsilon}}\frac{\left(
m^2 \right)^\epsilon}{\epsilon} + (3+2\epsilon)I_1(q^2=m^2) + 4m^2I_1^
\prime(q^2=m^2) \biggr) \biggr](\not q-m)^{-1}\not e^*_{A^\prime}
\biggr)u_A
\end{equation}
which restores the correct pole (and only pole) singularity of the complete
amplitude~Fig.~1 due to the quark $t$- channel intermediate state.

Now we sum the Eqs.~(\ref{216}) and~(\ref{32}) and obtain the complete
Regge
asymptotics ${\cal A} = {\cal A}_1 + {\cal A}_2$ which is to be compared
with the one-loop expansion of the Eq.~(\ref{12}) in order to find the
quark-gluon-Reggeized quark NLA effective vertices we are interested in.
This gives
$$
\Gamma_A(q,s_0) = gt^{A^\prime}\biggl( \not e^*_{A^\prime} + g^2C_F\biggl\{
\biggl[ \frac{\Gamma(1-\epsilon)}{(4\pi)^{2+\epsilon}}\frac{\sum_f\left(
m_f^2 \right)^\epsilon}{3\epsilon C_F} + \biggl( 1 + \frac{1}{2C_AC_F}
\biggr)\biggl( \frac{\Gamma(1-\epsilon)}{(4\pi)^{2+\epsilon}}\frac{\left(
m^2 \right)^\epsilon}{\epsilon}
$$
$$
+ (3+2\epsilon)I_1(q^2=m^2) \biggr) \biggr]\not e^*_{A^\prime} + \biggl(
3 + \frac{1}{C_AC_F} \biggr)2m^2I_1^\prime(q^2=m^2)\not e^*_{A^\prime} -
\biggl[ m(\not q-m)^{-1}\not e^*_{A^\prime}
$$
$$
+ \left( 2 - (1+\epsilon)\frac{(\not q-m)\not q}{2q^2} \right)\not e^*_
{A^\prime} - \left( 2m + (1+\epsilon)\frac{m^2-q^2}{q^2}\not q + (1+2
\epsilon)(\not q-m) \right)
$$
$$
\times\not e^*_{A^\prime}(\not q+m)^{-1} - \left( 1 + \epsilon - \not q
(\not q+m)^{-1} \right)\frac{\not q\not e^*_{A^\prime}(\not q+m)^{-1}}
{C_AC_F} \biggr]\left( I_1 - I_1(q^2=m^2) \right)
$$
$$
+ \frac{\not q-m}{C_AC_F}\biggl[ \frac{m^2-q^2}{2q^2}\not q\not e^*_
{A^\prime} - m\left( (1-\epsilon)m-\epsilon\not q \right)\not e^*_{A
^\prime}(\not q+m)^{-1} \biggr]I_2
$$
$$
+ \left( 2 + \frac{1}{C_AC_F} \right)\frac{m^2-q^2}{2q^2}(\not q-m)\not
q\not e^*_{A^\prime}I_3
$$
\begin{equation}\label{33}
+ \frac{m^2-q^2}{8q^2}(\not q+m)\not q\not e^*_{A^\prime}\biggl[ \left(
2 + \frac{1}{C_AC_F} \right)^2I_4 + \frac{1}{C_A^2C_F^2}I_5 - \frac{2}
{C_AC_F}\left( 2 + \frac{1}{C_AC_F} \right)I_6 \biggr] \biggr\} \biggl)u_A
\end{equation}
and
$$
\bar\Gamma_B(q,s_0) = \bar v_Bgt^{B^\prime}\biggl( \not e^*_{B^\prime}
+ g^2C_F\biggl\{ \not e^*_{B^\prime}\biggl[ \frac{\Gamma(1-\epsilon)}
{(4\pi)^{2+\epsilon}}\frac{\sum_f\left( m_f^2 \right)^\epsilon}{3\epsilon
C_F} + \biggl( 1 + \frac{1}{2C_AC_F} \biggr)\biggl( \frac{\Gamma(1-
\epsilon)}{(4\pi)^{2+\epsilon}}\frac{\left( m^2 \right)^\epsilon}{\epsilon}
$$
$$
+ (3+2\epsilon)I_1(q^2=m^2) \biggr) \biggr] + \not e^*_{B^\prime}\biggl(
3 + \frac{1}{C_AC_F} \biggr)2m^2I_1^\prime(q^2=m^2) - \biggl[ m\not e^*_
{B^\prime}(\not q-m)^{-1} + \not e^*_{B^\prime}\times
$$
$$
\left( 2 - (1+\epsilon)\frac{\not q(\not q-m)}{2q^2} \right) - (\not q+m)
^{-1}\not e^*_{B^\prime}\left( 2m + (1+\epsilon)\frac{m^2-q^2}{q^2}\not q
+ (1+2\epsilon)(\not q-m) \right)
$$
$$
- \frac{(\not q+m)^{-1}\not e^*_{B^\prime}\not q}{C_AC_F}\left( 1 +
\epsilon - (\not q+m)^{-1}\not q \right) \biggr]\left( I_1 - I_1(q^2=m^2)
\right) + \biggl[ \not e^*_{B^\prime}\not q\frac{m^2-q^2}{2q^2}
$$
$$
- m(\not q+m)^{-1}\not e^*_{B^\prime}\left( (1-\epsilon)m-\epsilon\not q
\right) \biggr]\frac{\not q-m}{C_AC_F}I_2 + \left( 2 + \frac{1}{C_AC_F}
\right)\frac{m^2-q^2}{2q^2}\not e^*_{B^\prime}\not q(\not q-m)I_3
$$
\begin{equation}\label{34}
+ \not e^*_{B^\prime}\not q(\not q+m)\frac{m^2-q^2}{8q^2}\biggl[ \left( 2
+ \frac{1}{C_AC_F} \right)^2I_4 + \frac{1}{C_A^2C_F^2}I_5 - \frac{2}{C_A
C_F}\left( 2 + \frac{1}{C_AC_F} \right)I_6 \biggr] \biggr\} \biggr)~.
\end{equation}

The vertices~(\ref{33}) and~(\ref{34}) (of course, they are related each to
other by an evident symmetry), together with the list of integrals of
the Appendix are the results of this paper.

\section{Discussion}
\setcounter{equation}{0}

We have calculated here the NLA quark-gluon-Reggeized quark effective
interaction vertices for the case of QCD with massive quarks. They are
to be applied for the determination of two-loop Regge trajectory of
massive Reggeized quark, for instance. For massless QCD such vertices
are known from Ref.~\cite{9} and we have checked that our result is in
complete agreement with that of the Ref.~\cite{9}. Let us note that in
the massless case in our approach one does not need to consider the
pole contribution at all - it is completely given by the Born amplitude
and the radiative corrections can contribute to the branch point
singularity only.

We also note that throughout all this paper we worked with the bare
coupling $g$ instead of the renormalized one $g_\mu$. In order to
remove the ultraviolet divergences from our results~(\ref{33})
and~(\ref{34}) it is enough to re-express them in terms of the renormalized
coupling (in $\overline{\mbox{MS}}$)
\begin{equation}\label{41}
g = g_\mu\mu^{-\epsilon}\left[ 1 + \left( \frac{11}{3} - \frac{2n_f}{3N}
\right)\frac{g_\mu^2N\Gamma(1-\epsilon)}{2\epsilon(4\pi)^{2+\epsilon}}
\right]~.
\end{equation}
At the moment such renormalization does not look so sensible to perform
since our results are in any case intermediate and contain infrared
poles in $\epsilon$ which would cancel only in final physical results.
As for the quark mass which enters the results, in our approach it
appears automatically as the renormalized pole quark mass because
we always used the unitarity relations with the renormalized
intermediate amplitudes. Of course, it is very easy to express back
through the bare mass (or through the mass in any other scheme).

\vspace{0.5cm}
{\bf\underline{Acknowledgement.}} M.~Kotsky thanks Dipartimento di Fisica,
Universit\'a della Calabria (Italy) for their warm hospitality while a
part of this work was done. A.~Principe thanks Dr. A.~Papa for his
kind attention to this work and many fruitful discussions.

\appendix
\section{Appendix}
\setcounter{equation}{0}

Here we present the results of integration in the Eqs.~(\ref{217}). Since
in the massive quark case the integrals can not be explicitly calculated
without the $\epsilon$- expansion we perform only those integrations
over Feynman parameters which can be done exactly and leave the others
untouched.
\begin{equation}\label{a1}
I_1 = -
\frac{\Gamma(1-\epsilon)}{(4\pi)^{2+\epsilon}}\frac{1}{\epsilon}
\int_0^1dx\left[ (1-x)\left( m^2 - xq^2 \right) \right]^\epsilon~,
\end{equation}
\begin{equation}\label{a2}
I_1(q^2=m^2) = -
\frac{\Gamma(1-\epsilon)}{(4\pi)^{2+\epsilon}}\frac{m^{2\epsilon}}
{\epsilon(1+2\epsilon)}~,
\end{equation}
\begin{equation}\label{a3}
I_1^\prime(q^2=m^2) = \frac{\Gamma(1-\epsilon)}{(4\pi)^{2+\epsilon}}
\frac{\left( m^2 \right)^{\epsilon-1}}{2\epsilon(1+2\epsilon)}~,
\end{equation}
\begin{equation}\label{a4}
I_2 = -
\frac{\Gamma(1-\epsilon)}{(4\pi)^{2+\epsilon}}\frac{1}{\epsilon
\left( m^2-q^2 \right)}\int_0^1\frac{dx}{x}\left( \left[
(1-x)\left( m^2 - xq^2 \right) \right]^\epsilon - \left[ (1-x)m
\right]^{2\epsilon} \right)~,
\end{equation}
\begin{equation}\label{a5}
I_3 =
-\frac{\Gamma(1-\epsilon)}{(4\pi)^{2+\epsilon}}\frac{1}{2\epsilon}
\int_0^1\frac{dx}{\left[ (1-x)\left( m^2 - xq^2 \right)
\right]^{1-\epsilon}}~,
\end{equation}
$$
I_4= -\frac{\Gamma(1-\epsilon)}{(4\pi)^{2+\epsilon}}
\int_0^1\frac{dx}{\left[ (1-x)\left( m^2 - xq^2 \right)
\right]^{1-\epsilon}}\left[ \frac{1}{\epsilon} + \psi(1) +
\psi(1-\epsilon) - 2\psi(1+2\epsilon) + \right.
$$
$$
\left.\ln\left( \frac{s_0}{(1-x)\left( m^2 - xq^2 \right)} \right)
\right] = -\frac{\Gamma(1-\epsilon)}{(4\pi)^{2+\epsilon}} \left[
\frac{1}{\epsilon} + \psi(1) + \psi(1-\epsilon) -
2\psi(1+2\epsilon) -
s_0^\epsilon\frac{d}{d\epsilon}s_0^{-\epsilon} \right]
$$
\begin{equation}\label{a6}
\times\int_0^1\frac{dx}{\left[ (1-x)\left( m^2 - xq^2 \right)
\right]^{1-\epsilon}}~,
\end{equation}
\begin{equation}\label{a7}
I_5 = I_4 + \frac{2\Gamma(1-\epsilon)}{(4\pi)^{2+\epsilon}}
\int_0^1\frac{dx\ln\left( x/(1-x) \right)}{\left[ (1-x)\left( m^2
- xq^2 \right) \right]^{1-\epsilon}}~,
\end{equation}
\begin{equation}\label{a8}
I_6 = I_4 + \frac{\Gamma(1-\epsilon)}{(4\pi)^{2+\epsilon}}
\int_0^1\frac{dx\ln\left( x/(1-x) \right)}{\left[ (1-x)\left( m^2
- xq^2 \right) \right]^{1-\epsilon}}~,
\end{equation}
where $\psi(z)$ is the logarithmic derivative of the Euler gamma-
function.

Let us note that the relation (see Eqs.~(\ref{a6}-\ref{a8}))
\begin{equation}\label{a9}
I_4 + I_5 - 2I_6 = 0~,
\end{equation}
valid for the Regge asymptotics of the above box integrals, is not
accidental. The matter is that such the combination of the boxes
is proportional to the purely nonasymptotic contribution (compare
with the Eq.~(\ref{21}))
\begin{equation}\label{a10}
{\cal A}_{na}^{(3,+)} = \left(
\int\frac{d^Dp}{i(2\pi)^D}\frac{\sum(-)B_{na}A_{na}}{\left( k^2 +
i\delta \right)\left( p^2 - m^2 + i\delta \right)} \right)^{(3,+)}
\end{equation}
to the Regge asymptotics of the amplitude ${\cal A}$ with quark
colour quantum numbers and positive signature in the $t$- channel.
In the above relation we have introduced $A_{na} = A - A_{as}$,
where the amplitude $A$ and its Regge asymptotics $A_{as}$ are
given by the Eqs.~(\ref{22})~and~(\ref{26}) respectively, and
analogously for the $B_{na}$. From other side, one can show
without real calculations that for the quark colour quantum
numbers in the $t$- channel the "$na\times na$" term can
contribute to the negative signature only and therefore
both~(\ref{a10})~and~(\ref{a9}) must vanish (see also
Ref.~\cite{12}).


\begin{thebibliography}{99}

\bibitem{1}
M.~Gell-Mann, M.L.~Goldberger, F.E.~Low, E.~Marx,
F.~Zachariasen, Phys. Rev. {\bf 133B} (1964) 145.

\bibitem{2}
S.~Mandelstam, Phys. Rev. {\bf 137B} (1964) 949.

\bibitem{3}
M.T.~Grisaru, H.J.~Schnitzer, H-S~Tsao, Phys. Lett.
{\bf 30} (1973) 811; Phys. Rev. {\bf D8} (1973) 4498;
L.N.~Lipatov, Yad. Fiz. {\bf 23} (1976) 642 [Sov. J. Nucl.
Phys. {\bf 23} (1976) 338].

\bibitem{4}
V.S.~Fadin, V.E.~Sherman, ZhETF Pis'ma {\bf 23} (1976) 599
[JETP Lett. {\bf 23} (1976) 548]; ZhETF {\bf 72} (1977) 1640 [Sov.
Phys. JETP {\bf 45(5)} (1977) 861].

\bibitem{5}
V.S.~Fadin, E.A.~Kuraev, L.N.~Lipatov, Phys. Lett.
{\bf B60} (1975) 50; E.A.~Kuraev, L.N.~Lipatov,
V.S.~Fadin, Zh. Eksp. Teor. Fiz. {\bf 71} (1976) 840
[Sov. Phys. JETP {\bf 44} (1976) 443]; {\bf 72} (1977)
377 [{\bf 45} (1977) 199]; Ya.Ya.~Balitskii, L.N.~Lipatov,
Sov. J. Nucl. Phys. {\bf 28} (1978) 822.

\bibitem{6}
V.S.~Fadin, L.N.~Lipatov, Zh. Eksp. Teor. Fiz. Pis'ma
{\bf 49} (1989) 311 [Sov. Phys. JETP Lett. {\bf 49}
(1989) 352]; Yad. Fiz. {\bf 50} (1989) 1141 [Sov. J.
Nucl. Phys. {\bf 50} (1989) 712]; Nucl. Phys. {\bf B477}
(1996) 767; V.S.~Fadin, R.~Fiore, A.~Quartarolo, Phys. Rev.
{\bf D50} (1994) 2265; 5893; {\bf D53} (1996) 2729;
V.S.~Fadin, R.~Fiore, M.I.~Kotsky, Phys. Lett. {\bf B359}
(1995) 181; {\bf B387} (1996) 593; {\bf B389} (1996) 737;
V.S.~Fadin, R.~Fiore, Phys. Lett. {\bf B294} (1992) 286;
V.S.~Fadin, Zh. Eksp. Teor. Fiz. Pis'ma {\bf 61} (1995) 342;
V.S.~Fadin, M.I.~Kotsky, Yad. Fiz. {\bf 59}(6) (1996) 1;
V.S.~Fadin, M.I.~Kotsky, L.N.~Lipatov, Phys. Lett. {\bf B415}
(1997) 97; Yad. Fiz. {\bf 61}(6) (1998) 716; S.~Catani,
M.~Ciafaloni, F.~Hautmann, Phys. Lett. {\bf B242} (1990) 97;
Nucl. Phys. {\bf B366} (1991) 135; G.~Camici, M.~Ciafaloni,
Phys. Lett. {\bf B386} (1996) 341; Nucl. Phys. {\bf B496}
(1997) 305; V.S.~Fadin, R.~Fiore, A.~Flachi, M.I.~Kotsky,
Phys. Lett. {\bf B422} (1998) 287; Yad. Fiz. {\bf 62}(6) (1999) 1.

\bibitem{7}
V.S.~Fadin, L.N.~Lipatov, Phys. Lett. {\bf B429} (1998) 127-134.

\bibitem{8}
M.~Ciafaloni, G.~Camici, Phys. Lett. {\bf B430} (1998) 349-354.

\bibitem{9}
V.S.~Fadin, R.~Fiore, Phys. Rev. {\bf D64} (2001) 114012.

\bibitem{10}
A.V.~Bogdan, V.~Del~Duca, V.S.~Fadin, E.W.N.~Glover, {\tt
hep-ph/0201240}.

\bibitem{11}
L.N.~Lipatov, M.I.~Vyazovsky, Nucl. Phys. {\bf B597} (2001) 399-409.

\bibitem{12}
V.S.~Fadin, L.N.~Lipatov, Nucl. Phys. {\bf B406} (1993) 259.

\end{thebibliography}
\end{document}